\documentclass[sigconf]{acmart}

\usepackage{subfig}

\renewcommand\footnotetextcopyrightpermission[1]{} % removes footnote with conference information in first column

\AtBeginDocument{%
  \providecommand\BibTeX{{%
    \normalfont B\kern-0.5em{\scshape i\kern-0.25em b}\kern-0.8em\TeX}}}

\setcopyright{none}
\begin{document}

% Title
\title{Design and Development of an Automated Coimagination Support System}

% Authors
\author{John Noel Victorino}
\affiliation{%
  \institution{Graduate School of Life Science \\ \& Systems Engineering\\
Kyushu Institute of Technology}
  \city{Kitakyushu}
  \country{Japan}
  \postcode{808-0196}
}
\email{victorino.john-noel783@mail.kyutech.jp}

\author{Naoto Fukunaga}
\affiliation{%
  \institution{Graduate School of Life Science \\ \& Systems Engineering\\
Kyushu Institute of Technology}
  \city{Kitakyushu}
  \country{Japan}
  \postcode{808-0196}
}
\email{fukunaga.naoto159@mail.kyutech.jp}

\author{Tomohiro Shibata}
\affiliation{%
  \institution{Graduate School of Life Science \\ \& Systems Engineering\\
Kyushu Institute of Technology}
  \city{Kitakyushu}
  \country{Japan}
  \postcode{808-0196}
}
\email{tom@brain.kyutech.ac.jp}

\renewcommand{\shortauthors}{Victorino and Shibata}

% Abstract
\begin{abstract}
Coimagination method is a novel approach to support interactive communication for activating three (3) cognitive functions: episodic memory, division of attention, and planning. These cognitive functions are known to decline at an early stage of mild cognitive impairment (MCI). In previous studies about the coimagination method, experimenters tested different settings in different care institutions. Out of these experiments, various measures were introduced, analyzed, and presented. However, ease of changing configuration based on participants, and a quick assessment of captured data remained challenging. Also, several observers and measurers are needed to conduct the coimagination method. In this paper, we propose the initial design and development of an automated coimagination support system that can handle such challenges. We aim to have an automated coimagination support system that can be used easily either by healthy participants or elderly participants via a natural voice interface. In this paper, our focus is to measure how well our proposed features work with elderly participants. Preliminary experiments were conducted with healthy participants, and notably, with actual elder participants. Healthy participants experienced longer speaking round and question-and-answer round than with elderly participants; while, the latter had preparation time before the speaking round. In these preliminary experiments, our initial system showed the capability to handle different configurations. Healthy participants have operated the system using voice, while elderly participants managed to use the system with minimal assistance.
\end{abstract}




\maketitle
\pagestyle{plain} % removes running headers

\section{Introduction}
Coimagination method has been proposed as a novel method to support interactive communication for activating three (3) cognitive functions: episodic memory, division of attention, and planning function, which decline at an early stage of mild cognitive impairment (MCI) \cite{Otake2009Initial}. It is hypothesized that the activation of these cognitive functions is effective for the prevention of dementia \cite{Otake2009Initial, Otake2009WithReacts, Otake2011CIMM}. Also, the generation of social networks among individuals helps in protecting individuals against dementia \cite{Otake2009Initial}. In the past, there have been numerous experiments with the coimagination method in care institutions in Japan, from the typical coimagination method to modified versions. In these studies, several observers and measurers were needed to facilitate and to capture and record data during the coimagination method, which were difficult at times. To address these challenges, we propose an initial design and development of an automated coimagination support system. We aim to have an automated coimagination support system that can be used easily by either healthy participants or elderly participants by voice. Specifically, in this paper, we want to test our initial proposed key features.

\begin{enumerate}
    \item The automated support system can accept participants' registration.
    \item The automated support system can search images for the participants using Google Custom Search API. The image will be used as a visual aid during the participant's speaking round.
    \item The automated support system can record audio during the session. This can be used to analyze the session afterward.
    \item The automated support system can be used with a voice interface.
\end{enumerate}

In this paper, we want to conduct a preliminary performance study measuring the proposed key features. We will test first with healthy participants to fine-tune our automated system before we conduct a preliminary performance study with elder participants, who will be the actual end-users of our automated coimagination support system.

\section{Coimagination Method}
\subsection{Key Principles}
This section describes the principles followed by the coimagination method, procedures, and variations done in previous studies.

Coimagination method successfully supports interactive communication and generation of social networks by following certain principles and strategies \cite{Otake2009Initial, Otake2009WithReacts}. 

\begin{enumerate}
    \item \raggedright Coimagination method should support interactive communication with images to activate three (3) cognitive functions \cite{Otake2009Initial, Otake2009WithReacts}.
    \item \raggedright Coimagination method should contribute to the generation of social networks among subjects through communication \cite{Otake2009Initial}.
    \item \raggedright Coimagination method should evoke feelings and thoughts, not just memories \cite{Otake2009WithReacts}.
    \item \raggedright Coimagination method should give equal time for each participant to talk about their topic, to ask and answer questions from other participants, and to give comments and reactions \cite{Otake2009WithReacts}.
    \item \raggedright Coimagination method should have measures for effectiveness \cite{Otake2009Initial}.
\end{enumerate} 

\subsection{Typical Coimagination Method Procedures}
A typical coimagination program is designed as follows \cite{Otake2009Initial, Otake2009WithReacts}.

\begin{itemize}
    \item \raggedright Coimagination program consists of five (5) sessions. Each session is held for about one (1) hour per week.
    \item \raggedright Each session has a unique, predetermined theme. Themes can include but not limited to ``favorite food'', ``favorite things'', ``health and food''.
    \item \raggedright The average number of participants is six (6).
    \item \raggedright Each participant brings three (3) images, on the average.
    \item \raggedright A session has two (2) rounds. The first round is for brief speech, where each participant talks about their topics using the images. The second round is for questions and answers from other participants to each other.
    \item \raggedright The fifth session is for memory tasks. Images presented throughout the four (4) sessions are displayed randomly to each participant. The participants guess the owner of the image and the theme of the image.
\end{itemize}

\subsection{Previous Implementations of Coimagination Method}
Coimagination method has been conducted in the past with different configurations and different analysis. There were programs held with older adults with cognitive decline, compared with a typical program held with healthy older adults \cite{Tsukawaki2011}. In this scenario, the number of participants changed from six (6) to four (4) \cite{Tsukawaki2011}. Time for each round and the number of images had been adjusted based on the participants' cognitive function, physical fitness, and concentration \cite{Tsukawaki2011}. Furthermore, the question-and-answer round was held immediately after the brief speech round so as not to forget \cite{Tsukawaki2011}. There were also programs held in multiple facilities from different Japan prefectures \cite{Otake2012WebBased}. A personal coimagination support system for a single participant was also explored to address elders, not in care home facilities and lack social interaction among other elders \cite{Otaki2017PersonalCoimagination}. Most importantly, different measures were being analyzed and evaluated, in addition to analyzing the results of the memory tasks. For instance, the number of comments which evoked reactions, such as laughter and wonder, were counted \cite{Otake2009WithReacts}. The intensity of cognitive activities was also measured using social network analysis \cite{Otake2011CIMM}. Conversation sessions were transcribed to analyze different aspects such as common topics through grouping to examine knowledge creation \cite{Otake2009KnowledgeCreativeActivity} and duplication of certain words and phrases by each participant to reflect the quality of interaction \cite{Otake2013DuplicationAnalysis}. Different audio features were also analyzed such as overlaps during conversation  \cite{Otake2013HARK, Otake2013OverlapAnalysis} and detection of laughter \cite{Nergui2013LaughterDetection, Nergui2013LaughterDetectionUsingRespiratorySensor} to assist in group conversations.

\subsection{Motivation for the Proposed Automated Support System}
Given the context in the previous subsection, the coimagination method can be configured and analyzed differently, yet we want to make sure that the principles of the coimagination method are kept intact. We want to provide an automated tool for facilitators, observers, and measurers to easily conduct the coimagination method and analyze the different facets of the method. We intend that the automated support system will help facilitators connect more with the participants rather than handling administrative tasks in conducting the coimagination method. Moreover, we want the automated support system to be deployed in nursing care facilities and can be easily used by elders. In this way, we keep them engaged among themselves using the framework set by the coimagination method.

\section{Automated Coimagination Support System}
\subsection{System Architecture}
\begin{figure}[h]
\centering
\includegraphics[width=3.125in]{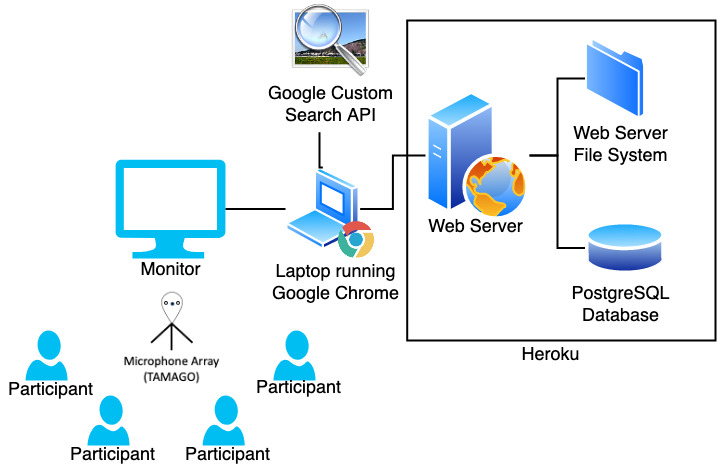}
\caption{The Automated Coimagination Support System Setup}
\label{fig:system_setup}
\end{figure}

The initial automated coimagination support system is mainly built using Web technologies. Ruby on Rails (RoR) is used as the primary web framework, with PostgreSQL for the database. Audio recordings were stored in the server's file system and not in the database. For speech processing, Web Speech (Application Programming Interface) API and Web Audio API were used to recognize speech commands and to record audio, respectively. Google Custom Search API is used to search images for participants, given specific keywords. The web application is deployed to Heroku, a cloud platform-as-a-service (PaaS). During the coimagination session, any laptop connected to the Internet and with the latest Google Chrome browser can be used to run the automated system. An external monitor is used for a bigger display. We also used an 8-channel microphone array for the audio input, but an internal laptop microphone can also be used. Figure \ref{fig:system_setup} shows an overview of the system architecture.

\subsection{User Interface}
In this section, the automated coimagination support system is presented with user interface screens. The automated system can be operated by voice commands via Web Speech API and microphone, or by clicking using the mouse. The automated coimagination support system is used as follows.

% \begin{figure*}[!t]
% \centering
% \subfloat[Home Menu]{\includegraphics[width=.3\textwidth]{1.jpg}%
% \label{fig:home_menu}}
% \hfil
% \subfloat[User Registration]{\includegraphics[width=.3\textwidth]{2.png}%
% \label{fig:user_registration}}
% \hfil
% \subfloat[Session Selection]{\includegraphics[width=.3\textwidth]{4.png}%
% \label{fig:session_selection}}
% \hfil
% \subfloat[Image Search]{\includegraphics[width=.3\textwidth]{5.jpg}%
% \label{fig:image_search}}
% \hfil
% \subfloat[Preparation Period]{\includegraphics[width=.3\textwidth]{6.png}%
% \label{fig:preparation_period}}
% \hfil
% \subfloat[Speaking Round]{\includegraphics[width=.3\textwidth]{7.jpg}%
% \label{fig:speaking_round}}
% \caption{User Interface Screens}
% \end{figure*}

\begin{enumerate}
    \item The initial home screen presents two options to register their names or start the session to the participants as seen in Figure \ref{fig:home_menu}.
    
    \begin{figure}[h]
    \centering
    \includegraphics[width=2.12in]{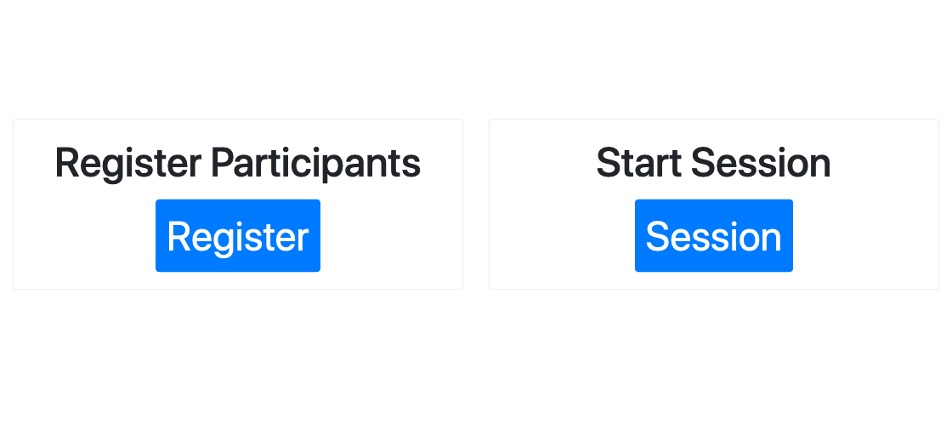}
    \caption{User Interface Screen: Home Menu}
    \label{fig:home_menu}
    \end{figure}
    
    \item Participants are asked for their names, as shown in Figure \ref{fig:user_registration}. The system will ask confirmation if it recognized the name correctly. The participant can say ``save'' or ``yes'' or click ``Save'' to confirm. In the Japanese version, ``hai'' is also accepted. After every participant has registered their names, saying or clicking ``back'' returns the screen to the home screen (Figure \ref{fig:home_menu}), where the session can be started by saying or clicking ``session''.
    
    \begin{figure}[h]
    \centering
    \includegraphics[width=2.12in]{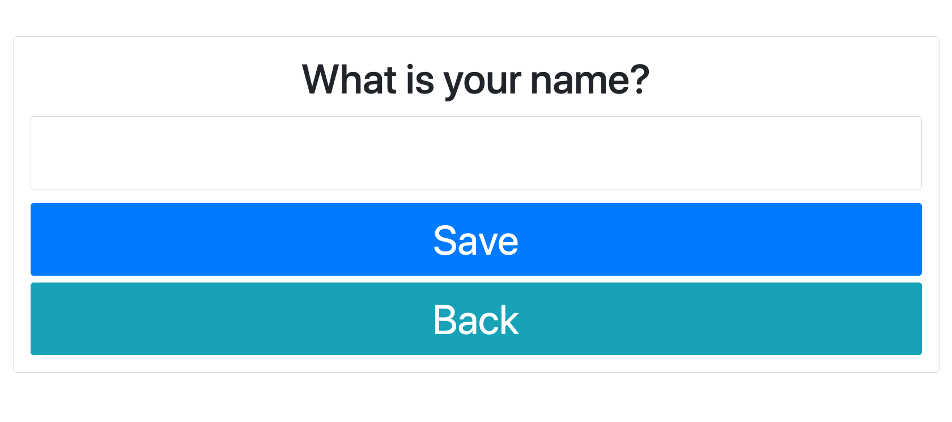}
    \caption{User Interface Screen: User Registration}
    \label{fig:user_registration}
    \end{figure}

    \item In Figure \ref{fig:session_selection}, an active session can be selected by saying the session's theme, for example, ``favorite food''.
    
    \begin{figure}[h]
    \centering
    \includegraphics[width=2.25in]{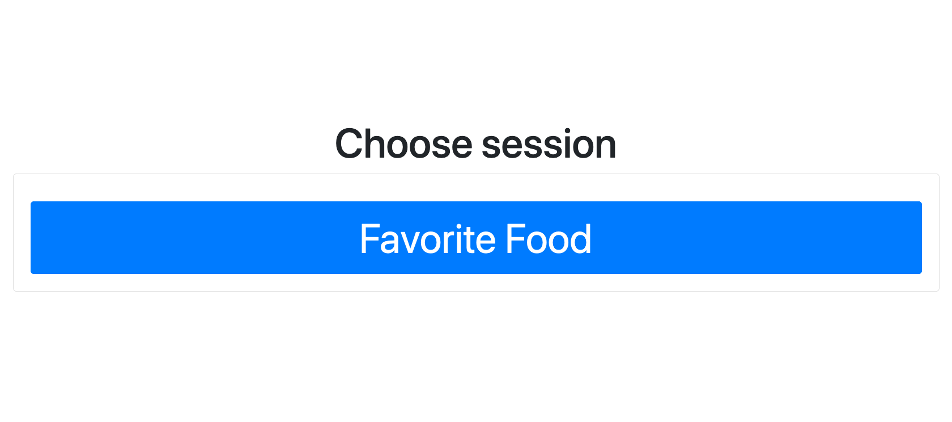}
    \caption{User Interface Screen: Session Selection}
    \label{fig:session_selection}
    \end{figure}
    
    \item Similar to step 2, each participant is asked for their response to the theme of the session. For example, the participant says ``fried chicken''. The system uses the keyword ``fried chicken`` as an input to Google Custom Search API. The system automatically searches images for ``fried chicken`` as if one searches using Google Search Engine. The system displays the top image result, as shown in Figure \ref{fig:image_search}.
    
    \begin{figure}[h]
    \centering
    \includegraphics[width=2.25in]{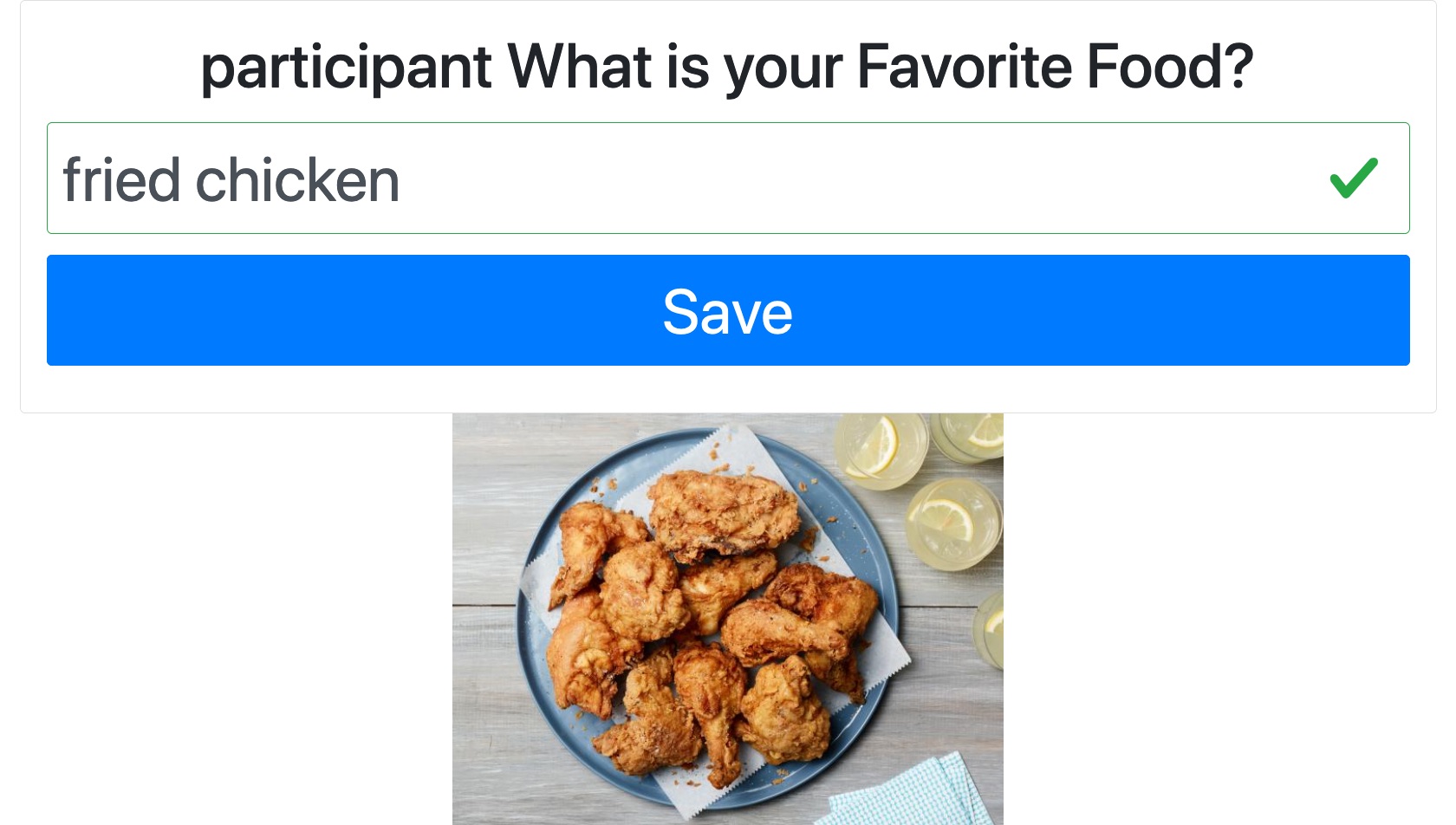}
    \caption{User Interface Screen: Image Search}
    \label{fig:image_search}
    \end{figure}
    
    \item To compensate for the lack of preparation time, each participant is given an arbitrary period of five (5) minutes to think about their topic. When the participant feels ready, the button can be pressed to proceed to the speaking round. This is seen in Figure \ref{fig:preparation_period}.
        
    \begin{figure}[h]
    \centering
    \includegraphics[width=2.25in]{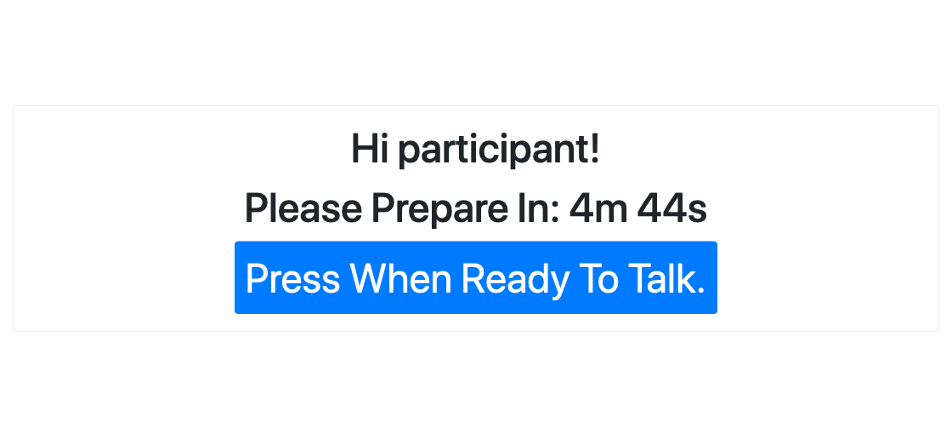}
    \caption{User Interface Screen: Preparation Period}
    \label{fig:preparation_period}
    \end{figure}
    
    \item Each participant is given one (1) minute thirty (30) seconds to speak. The enlarged image is shown on the left side; while, the timer, name of the participant speaking, the session's theme, and the participant's topic are shown on the right side, as shown in Figure \ref{fig:speaking_round}. This step repeats until the last participant is finished speaking, following the original coimagination method \cite{Otake2009WithReacts}.
    
    \begin{figure}[h]
    \centering
    \includegraphics[width=2.25in]{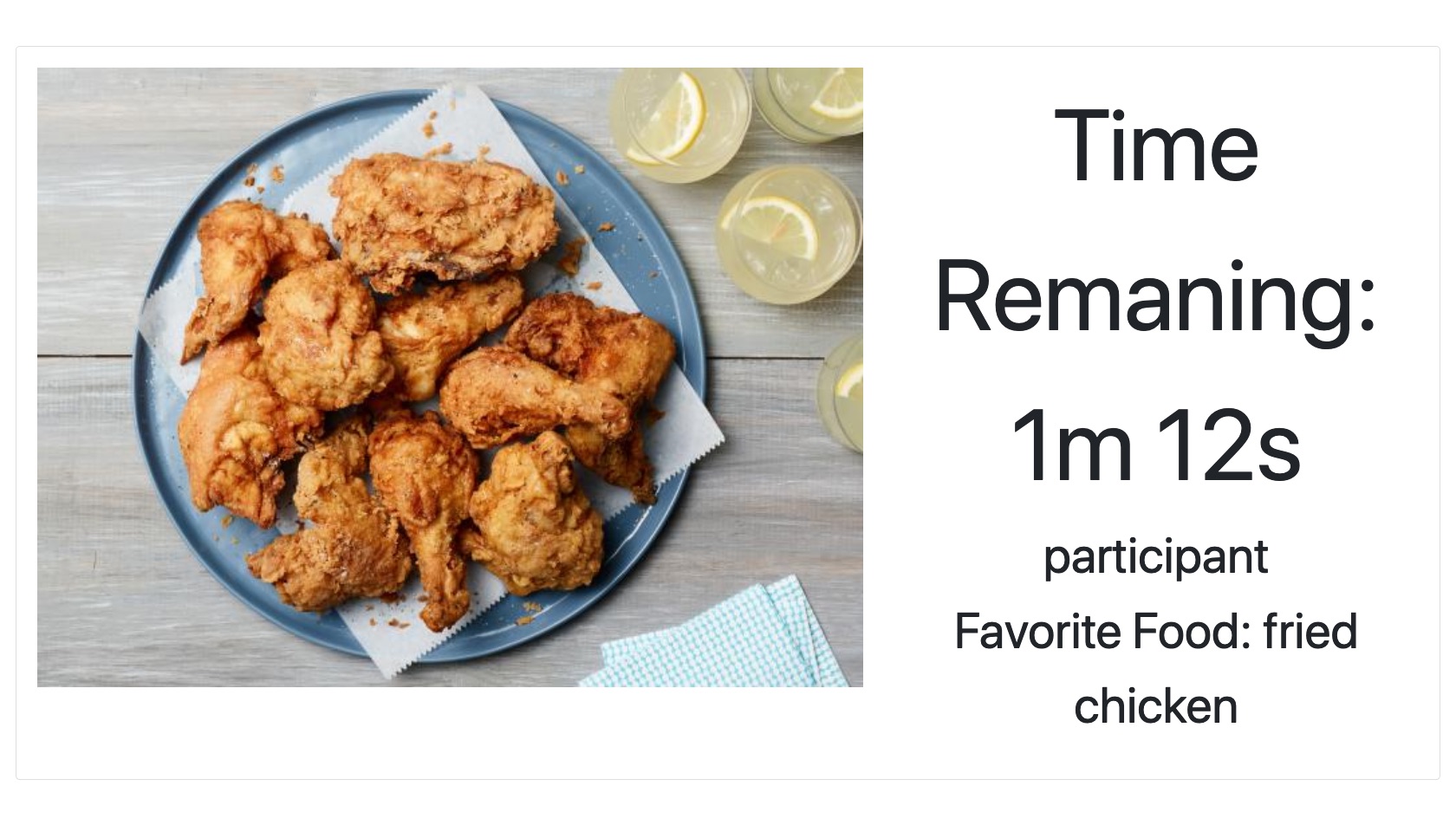}
    \caption{User Interface Screen: Speaking Round}
    \label{fig:speaking_round}
    \end{figure}
    
    \item Similar to step 5, participants are given a preparation period before the question-and-answer round. The only difference in this step is that no time limit is provided and the button has to be pressed to proceed.
    
    \item In this configuration, the question-and-answer round is given one (1) minute and thirty (30) seconds for each participant. The same screen setup is used in step 6.
    
    \item After the last participant has finished the question-and-answer round, the system shows a thank-you message.
\end{enumerate}

\subsection{System Configuration}
The automated coimagination support system can accept certain parameters to configure its user interface. The language can switch between English and Japanese. In effect, the labels and accepted spoken commands are adjusted based on the selected language. For example, the keyword ``hai'' is accepted in exchange for the keyword ``yes'' or keyword ``tōroku'' is accepted in exchange for the keyword ``register''. The themes for each session can also be added. The themes, as well as participants, can also be activated or deactivated to show or hide in the system during a session, respectively. The period for preparation, speaking round and question-and-answer can be changed easily using the system.

\section{Experiments}
\subsection{Experiments Setup}
A trial run (or alpha test) with healthy University students and a preliminary performance study were conducted to test the developed system. The trial run was conducted with four (4) healthy male University students, with an average age of 24. The speaking round and the question-and-answer round were given five (5) minutes. Preparation time for both rounds was not set because this feature was only added after the trial run. The addition of preparation time feature shall be explained in the next section. The trail run is meant to fine-tune the automated support system before the actual preliminary performance experiment with actual end-users.

\begin{figure}[h]
\centering
\includegraphics[width=2.25in]{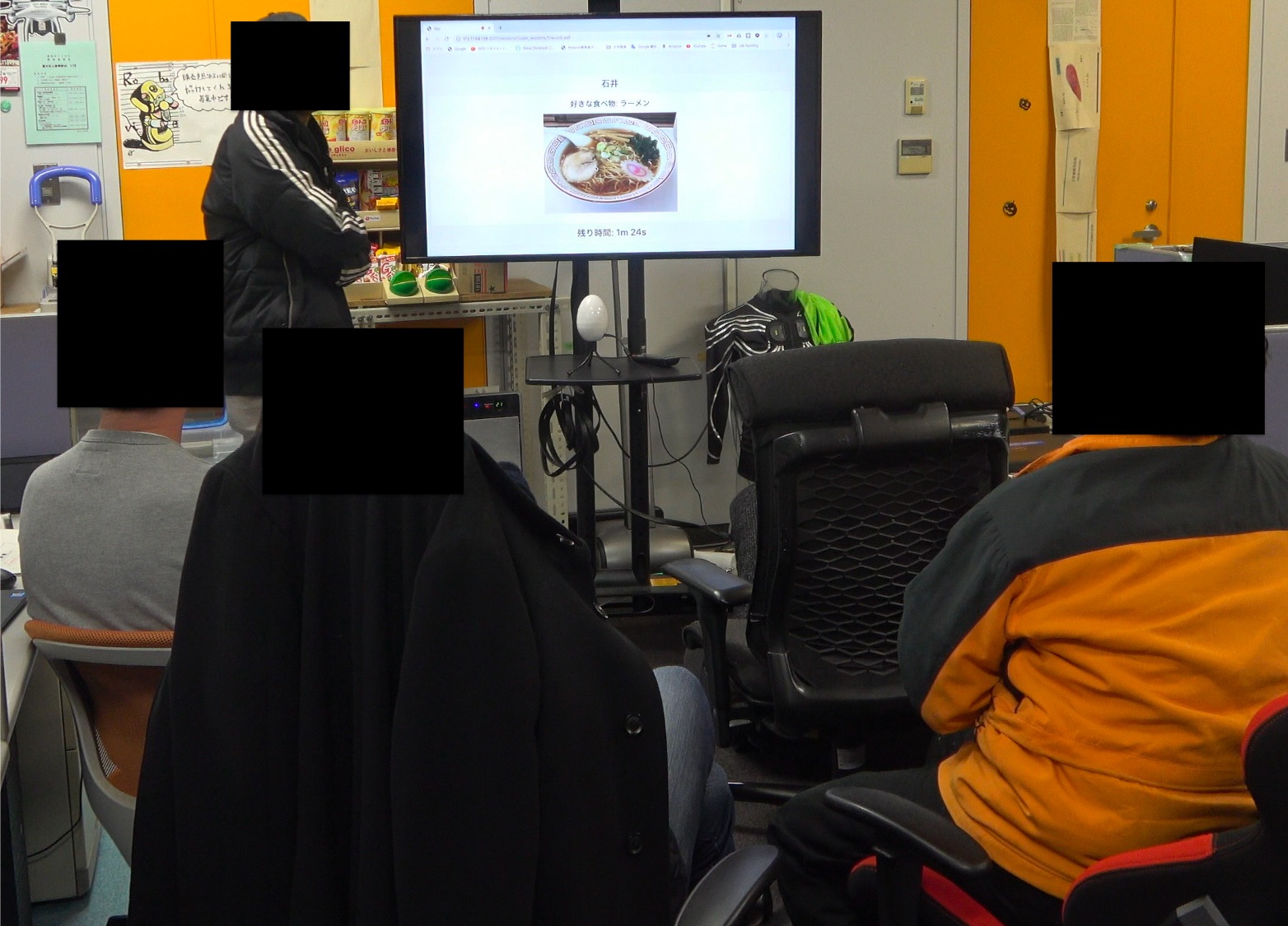}
\caption{Experiment with Healthy Participants}
\label{fig:healthy}
\end{figure}

\begin{figure}[h]
\centering
\includegraphics[width=2.25in]{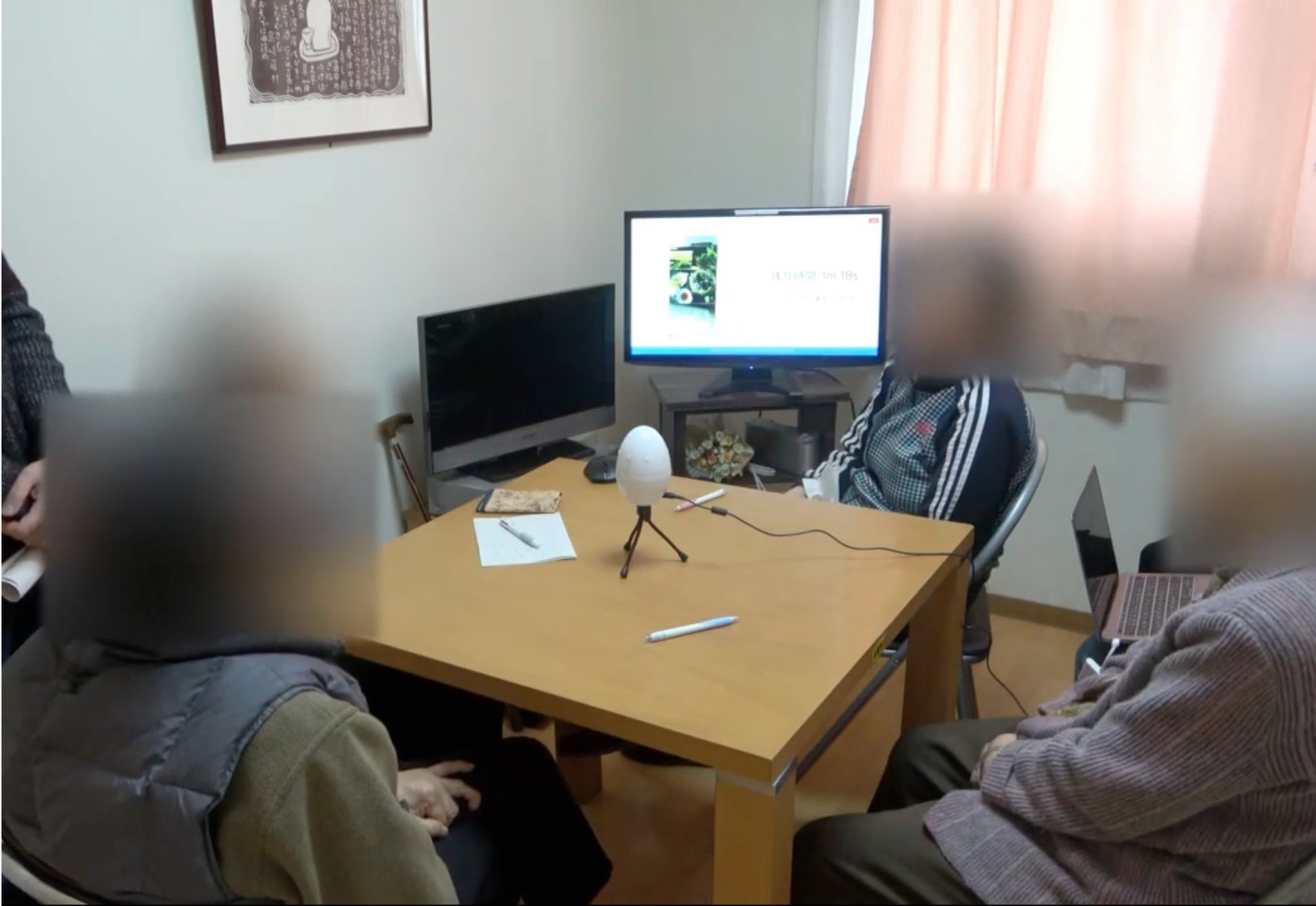}
\caption{Experiment with Elders Participants}
\label{fig:elders}
\end{figure}

The preliminary performance experiment (or beta test) was conducted at an elderly nursing care facility in Kitakyushu City, Fukuoka, Japan. There were four (4) elderly participants (one male, three female) with an average age of 84. All four elders have dementia but with varying severity. Speaking round and question-and-answer round were reduced to one (1) minute and thirty (30) seconds, with a preparation time of at most five (5) minutes before speaking round. The theme for both experiments is ``favorite food.'' Table \ref{table:experiments} summarizes the trial run and the preliminary experiment, while Figures \ref{fig:healthy} and \ref{fig:elders} show images with healthy participants and elderly participants, respectively.

\begin{table*}
  \caption{Experiments Setup Summary}
  \label{table:experiments}
  \begin{tabular}{lcc}
    \toprule
     & Trial Run & Preliminary Experiment \\
    \midrule
    Attribute & Healthy University students & Elders with varying severity of dementia \\
    Average Age (in years old) & 24 & 84 \\
    Gender Composition & 4 male & 1 male, 3 female \\
    Preparation Time before Speaking Round & Not applicable & 5 minutes \\
    Speaking round period & 5 minutes & 1 minute 30 seconds \\
    Question-\&-Answer round period & 5 minutes & 1 minute 30 seconds \\
    \bottomrule
  \end{tabular}
\end{table*}

\subsection{Experiments Measure}
This paper focuses on testing the performance of the proposed automated coimagination support system. Specifically, we want to measure the following proposed key features:

\begin{enumerate}
    \item The automated support system can accept participants' registration.
    \item The automated support system can search images for the participants using Google Custom Search API.
    \item The automated support system can record audio during the session.
    \item The automated support system can be used with a voice interface.
\end{enumerate}

Table \ref{tab:measures} shows the key features with the participants' input and expected output. It also shows the total number of attempts and the success rate. The total number of attempts is counted from video recordings during the preliminary experiment. Some participants only took one (1) attempt at a particular feature, while some participants took more than one (1) attempt. The success rate is computed based on the total successful attempts overall attempts.

\begin{table*}
  \caption{Preliminary Experiment Measures}
  \label{tab:measures}
  \begin{tabular}{p{3cm}p{2.5cm}p{2.5cm}rr}
    \toprule
    Key Features & Participant Input & Expected Output & Total number of attempts & Success Rate (\%) \\
    \midrule
    Participant Registration & Participant's name & System should record the name of the participants & 4 & 100.00\% \\
    Image Search & Participant's response to the session's theme & System should show the correct image. & 4 & 100.00\% \\
    Audio Recording: Speaking Round & Participant's voice during his/her discussion & System should record the participant's voice in an audio file during this round. & 4 & 100.00\% \\
    Audio Recording: Question-and-Answer Round & Participants' voices during their discussion & System should record the participants' voices in an audio file during this round. & 4 & 100.00\% \\
    Voice Interface: Participant Registration & Spoken participant's name & System should record the name of the participants & 8 & 0.00\% \\
    Voice Interface: Image Search & Spoken participant's response to the session's theme & System should show the correct image. & 9 & 67.50\% \\
    \bottomrule
  \end{tabular}
\end{table*}

\section{Discussion}
The primary purpose of the trial run with University students is to fine-tune the automated support system before the actual preliminary experiment with actual end-users, i.e., elders. The automated system has been used easily since the participants are University students who are at least familiar with voice-based applications. However, we noticed that five (5) minutes for both speaking and question-and-answer round was too long. Participants were neither talking for a certain amount of time nor talking about the theme of the session. Due to these observations, we decided to reduce the time for the second experiment with elders. In previous coimagination experiments, time can be adjusted based on the composition of the participants \cite{Tsukawaki2011}. Nonetheless, the system has successfully fulfilled its goals – registering and recording the coimagination session, supporting image preparation using Google Custom Search API, recording audio for both rounds, and navigating the system using the voice interface.

During the preliminary experiment, there were more observations to be considered. First, the elders responded to the system in complete and polite sentences, denoted by the use of ``I am ...'', and ``desu'' (which in the Japanese language denotes certain politeness). This is the reason why the system could not correctly register the participant's name by voice interface. After failing two (2) times, we decided to type their names instead. The input field has been added as a fail-safe feature.

The same happened when asked for their favorite food. They responded with a sentence pattern ``I like ...''. Two (2) out of the four (4) participants' voice-based response was properly recognized. Thus, image searching by voice has been accomplished. However, for the other two (2) participants, we assisted them in their response by voice instead of typing their responses. 

Another possible reason for the low success rate of the voice interface happened when the system could also hear other participants talking. The system could not determine the end of the sentence. Thus, the recognized input is a mixture of the recognized voice from two or more participants.

In the aspect of correctly providing an image based on certain input (i.e.,``What is your favorite food?''), the system successfully returned the correct image during the preliminary experiment. It happened during the development phase where we tested a few ``favorite food'', but the system did not return a correct intended image. For example, ``naruto'' is Japanese fish cake. But because of the popularity of the fictional character ``Naruto'', the system returned an incorrect image. Fortunately, during the preliminary experiment, the participants did not respond like this.

Regarding the changes between the trial run and the preliminary experiment, the elders still showed a period of silence during speaking round and question-and-answer round, despite the decrease in time. We, together with the nursing care facility staff, had to ask questions or make short comments so that participants continue to talk. Participants who are listeners during the speaking round were sometimes making comments. In general, the system seemed to engage the participants to communicate with each other.

\subsection{Comparison Between the Typical Coimagination Method and the Automated Coimagination Support System}
The automated coimagination support system was developed to closely follow the typical coimagination method, as much as possible. Changes have been made to address the previously mentioned challenges.

\begin{itemize}
    \item In a typical coimagination method, the participants are informed about the theme of the session. Thus, they can find and bring their own pictures to be presented during the session. In the automated method, the participants inform the system of their response to the theme. Then, the system finds a picture from the Internet using Google Custom Search API.
    \item In relation to the previous one, the automated method has a preparation period before the speaking round of each participant. The preparation preiod was added as a cautionary feature for the participants. They can collect their thoughts about the topic, remember the feelings and events related to their topic. The participant can proceed to the speaking round if they feel they are ready.
    \item In a typical coimagination method, a human facilitator operates a laptop with a large screen. Before the session, the facilitator scans the images into the laptop and arranges them according to seat order. During the session, the facilitator chooses the images based on whose turn it is, enlarges a specific image based on the participant's request, and can ask short questions to assist the participant in their topic. In other studies, additional observers were needed to measure certain things e.g., the number of comments or reactions during the session. In the automated method, we aim that the participants can operate the system by themselves. If the system fails to function properly, as it happened during the preliminary experiments, manual input via keyboard can be done by their supervisor. Moreover, the automated system assists the supervisor in data collection. In this way, listening and responding to the participants is the main purpose of the supervisor.
\end{itemize}

\section{Conclusion}
In this paper, we presented the initial design and development of an automated coimagination support system that can register and record coimagination sessions, search for images for the participants using Google Custom Search API, and record audio during the session and can be used with a voice interface.  The automated system can be configured easily in terms of language support, the period for each round and phase, and the addition, activation, and deactivation of sessions and users. We first tested the developed system with University students as a trial run. Then, we tested the re-configured system with actual end-users during the preliminary experiment. In this performance study, we observed and analyzed how well the system performed based on the preliminary experiment's video recording. We have successfully demonstrated three (3) out of the four (4) key features that we propose. Our future versions can include noise cancellation and better speech separation for improved voice recognition due to the voice interface challenge. Additionally, natural language processing (NLP) tools such as intent classification can be added to handle the complete sentences provided by the participants. We also plan to add multiple images per participant in a session and taking a picture of one's own images to handle the issue on the correct image.

\section{Acknowledgments}
The authors would like to thank the Ministry of Education, Culture, Sports, Science and Technology Japan (MEXT) for the scholarship grant and Sawayaka Club of Kitakyushu for allowing us to conduct our experiment. This work was supported by JSPS KAKENHI Grant Number JP16H06534.

\bibliographystyle{ACM-Reference-Format}
\bibliography{references}


\begin{thebibliography}{12}

%%% the Web address.

\ifx \showCODEN    \undefined \def \showCODEN     #1{\unskip}     \fi
\ifx \showDOI      \undefined \def \showDOI       #1{#1}\fi
\ifx \showISBNx    \undefined \def \showISBNx     #1{\unskip}     \fi
\ifx \showISBNxiii \undefined \def \showISBNxiii  #1{\unskip}     \fi
\ifx \showISSN     \undefined \def \showISSN      #1{\unskip}     \fi
\ifx \showLCCN     \undefined \def \showLCCN      #1{\unskip}     \fi
\ifx \shownote     \undefined \def \shownote      #1{#1}          \fi
\ifx \showarticletitle \undefined \def \showarticletitle #1{#1}   \fi
\ifx \showURL      \undefined \def \showURL       {\relax}        \fi
% invisible to TeX
\providecommand\bibfield[2]{#2}
\providecommand\bibinfo[2]{#2}
\providecommand\natexlab[1]{#1}
\providecommand\showeprint[2][]{arXiv:#2}

\bibitem[\protect\citeauthoryear{Nergui and Otake}{Nergui and Otake}{2013}]%
        {Nergui2013LaughterDetection}
\bibfield{author}{\bibinfo{person}{Myagmarbayar Nergui} {and}
  \bibinfo{person}{Mihoko Otake}.} \bibinfo{year}{2013}\natexlab{}.
\newblock \showarticletitle{Laughter detection for an assisting tool of group
  conversation}.
\newblock \bibinfo{journal}{\emph{methods}} \bibinfo{volume}{8},
  \bibinfo{number}{9} (\bibinfo{year}{2013}), \bibinfo{pages}{10}.
\newblock


\bibitem[\protect\citeauthoryear{Nergui, Pai-hui, Nagamatsu, Waki, and
  Otake}{Nergui et~al\mbox{.}}{2013}]%
        {Nergui2013LaughterDetectionUsingRespiratorySensor}
\bibfield{author}{\bibinfo{person}{Myagmarbayar Nergui}, \bibinfo{person}{Lin
  Pai-hui}, \bibinfo{person}{Gota Nagamatsu}, \bibinfo{person}{Mikio Waki},
  {and} \bibinfo{person}{Mihoko Otake}.} \bibinfo{year}{2013}\natexlab{}.
\newblock \showarticletitle{Automatic detection of laughter using respiratory
  sensor data with smile degree}. In \bibinfo{booktitle}{\emph{Proceedings of
  the International Conference on Advances in Computing, Communications and
  Informatics}}, Vol.~\bibinfo{volume}{1}. \bibinfo{pages}{8--11}.
\newblock


\bibitem[\protect\citeauthoryear{Otake}{Otake}{2009}]%
        {Otake2009WithReacts}
\bibfield{author}{\bibinfo{person}{Mihoko Otake}.}
  \bibinfo{year}{2009}\natexlab{}.
\newblock \showarticletitle{Coimagination method: sharing imagination with
  images and time limit}. In \bibinfo{booktitle}{\emph{Proceedings of the
  International Reminiscence and Life Review Conference}},
  Vol.~\bibinfo{volume}{2009}. \bibinfo{pages}{97--103}.
\newblock


\bibitem[\protect\citeauthoryear{Otake, Kato, Iwata, Takagi, and Asama}{Otake
  et~al\mbox{.}}{2009a}]%
        {Otake2009KnowledgeCreativeActivity}
\bibfield{author}{\bibinfo{person}{Mihoko Otake}, \bibinfo{person}{Motoichiro
  Kato}, \bibinfo{person}{Shuichi Iwata}, \bibinfo{person}{Toshihisa Takagi},
  {and} \bibinfo{person}{Hajime Asama}.} \bibinfo{year}{2009}\natexlab{a}.
\newblock \showarticletitle{Knowledge Creative Activity Support towards
  Prevention of Dementia via Coimagination Method}. In
  \bibinfo{booktitle}{\emph{The 23rd Annual Conference of the Japanese Society
  for Artificial Intelligence}}. The Japanese Society for Artificial
  Intelligence, \bibinfo{pages}{2F1NFC41--2F1NFC41}.
\newblock


\bibitem[\protect\citeauthoryear{Otake, Kato, Takagi, and Asama}{Otake
  et~al\mbox{.}}{2009b}]%
        {Otake2009Initial}
\bibfield{author}{\bibinfo{person}{Mihoko Otake}, \bibinfo{person}{Motoichiro
  Kato}, \bibinfo{person}{Toshihisa Takagi}, {and} \bibinfo{person}{Hajime
  Asama}.} \bibinfo{year}{2009}\natexlab{b}.
\newblock \showarticletitle{Coimagination method: Communication support system
  with collected images and its evaluation via memory task}. In
  \bibinfo{booktitle}{\emph{International Conference on Universal Access in
  Human-Computer Interaction}}. Springer, \bibinfo{pages}{403--411}.
\newblock


\bibitem[\protect\citeauthoryear{Otake, Kato, Takagi, and Asama}{Otake
  et~al\mbox{.}}{2011}]%
        {Otake2011CIMM}
\bibfield{author}{\bibinfo{person}{Mihoko Otake}, \bibinfo{person}{Motoichiro
  Kato}, \bibinfo{person}{Toshihisa Takagi}, {and} \bibinfo{person}{Hajime
  Asama}.} \bibinfo{year}{2011}\natexlab{}.
\newblock \showarticletitle{The coimagination method and its evaluation via the
  conversation interactivity measuring method}.
\newblock In \bibinfo{booktitle}{\emph{Early Detection and Rehabilitation
  Technologies for Dementia: Neuroscience and Biomedical Applications}}.
  \bibinfo{publisher}{IGI Global}, \bibinfo{pages}{356--364}.
\newblock


\bibitem[\protect\citeauthoryear{Otake, Koyanagi, Doi, Tsujihata, Tasaki,
  Noguchi, Abe, and Nagata}{Otake et~al\mbox{.}}{2012}]%
        {Otake2012WebBased}
\bibfield{author}{\bibinfo{person}{Mihoko Otake}, \bibinfo{person}{Yoko
  Koyanagi}, \bibinfo{person}{Yukie Doi}, \bibinfo{person}{Mitsuhiro
  Tsujihata}, \bibinfo{person}{Takayo Tasaki}, \bibinfo{person}{Muneaki
  Noguchi}, \bibinfo{person}{Akira Abe}, {and} \bibinfo{person}{Eiko Nagata}.}
  \bibinfo{year}{2012}\natexlab{}.
\newblock \showarticletitle{Development of Coimagination Method Support System
  named ``Fonobono Panel'' Utilized at Multiple Facilities}. In
  \bibinfo{booktitle}{\emph{The 26th Annual Conference of the Japanese Society
  for Artificial Intelligence}}. The Japanese Society for Artificial
  Intelligence, \bibinfo{pages}{2A1NFC613--2A1NFC613}.
\newblock


\bibitem[\protect\citeauthoryear{Otake, Nergui, Moon, Takagi, Kamashima, and
  Nakadai}{Otake et~al\mbox{.}}{2013a}]%
        {Otake2013HARK}
\bibfield{author}{\bibinfo{person}{Mihoko Otake}, \bibinfo{person}{Myagmarbayar
  Nergui}, \bibinfo{person}{Seong-eun Moon}, \bibinfo{person}{Kentaro Takagi},
  \bibinfo{person}{Tsutomu Kamashima}, {and} \bibinfo{person}{Kazuhiro
  Nakadai}.} \bibinfo{year}{2013}\natexlab{a}.
\newblock \showarticletitle{Development of a sound source localization system
  for assisting group conversation}. In \bibinfo{booktitle}{\emph{International
  Conference on Intelligent Robotics and Applications}}. Springer,
  \bibinfo{pages}{532--539}.
\newblock


\bibitem[\protect\citeauthoryear{Otake, Nergui, Otani, and Ota}{Otake
  et~al\mbox{.}}{2013b}]%
        {Otake2013DuplicationAnalysis}
\bibfield{author}{\bibinfo{person}{Mihoko Otake}, \bibinfo{person}{Myagmarbayar
  Nergui}, \bibinfo{person}{Takashi Otani}, {and} \bibinfo{person}{Jun Ota}.}
  \bibinfo{year}{2013}\natexlab{b}.
\newblock \showarticletitle{Duplication analysis of conversation and its
  application to cognitive training of older adults in care facilities}.
\newblock \bibinfo{journal}{\emph{Journal of Medical Imaging and Health
  Informatics}} \bibinfo{volume}{3}, \bibinfo{number}{4}
  (\bibinfo{year}{2013}), \bibinfo{pages}{615--621}.
\newblock


\bibitem[\protect\citeauthoryear{Otake and Yamaguchi}{Otake and
  Yamaguchi}{2013}]%
        {Otake2013OverlapAnalysis}
\bibfield{author}{\bibinfo{person}{Mihoko Otake} {and} \bibinfo{person}{Kenta
  Yamaguchi}.} \bibinfo{year}{2013}\natexlab{}.
\newblock \showarticletitle{Analysis of Overlap during Group Conversation of
  Active Older Adults}. In \bibinfo{booktitle}{\emph{The 27th Annual Conference
  of the Japanese Society for Artificial Intelligence}}. The Japanese Society
  for Artificial Intelligence, \bibinfo{pages}{3C1IOS1b5--3C1IOS1b5}.
\newblock


\bibitem[\protect\citeauthoryear{Otaki and Otake}{Otaki and Otake}{2017}]%
        {Otaki2017PersonalCoimagination}
\bibfield{author}{\bibinfo{person}{Hikaru Otaki} {and} \bibinfo{person}{Mihoko
  Otake}.} \bibinfo{year}{2017}\natexlab{}.
\newblock \showarticletitle{Interactive Robotic System Assisting Image Based
  Dialogue for the Purpose of Cognitive Training of Older Adults}. In
  \bibinfo{booktitle}{\emph{2017 AAAI Spring Symposium Series}}.
\newblock


\bibitem[\protect\citeauthoryear{Tsukawaki, Tadenuma, Sato, Negishi, Taguchi,
  Maekawa, Nagai, Takeshita, Kuroda, Myojin, Onitake, Hasegawa, and
  Otake}{Tsukawaki et~al\mbox{.}}{2011}]%
        {Tsukawaki2011}
\bibfield{author}{\bibinfo{person}{Akio Tsukawaki}, \bibinfo{person}{Yoshiyasu
  Tadenuma}, \bibinfo{person}{Yukiko Sato}, \bibinfo{person}{Katsutoshi
  Negishi}, \bibinfo{person}{Yoshie Taguchi}, \bibinfo{person}{Akiko Maekawa},
  \bibinfo{person}{Sumiko Nagai}, \bibinfo{person}{Hideko Takeshita},
  \bibinfo{person}{Seiji Kuroda}, \bibinfo{person}{Yoshiki Myojin},
  \bibinfo{person}{Makato Onitake}, \bibinfo{person}{Yoshinori Hasegawa}, {and}
  \bibinfo{person}{Mihoko Otake}.} \bibinfo{year}{2011}\natexlab{}.
\newblock \showarticletitle{Development of Support Service for Care Welfare
  Facilities based on Coimagination Method towards Prevention and Recovery from
  Dementia}. In \bibinfo{booktitle}{\emph{The 25th Annual Conference of the
  Japanese Society for Artificial Intelligence}}. The Japanese Society for
  Artificial Intelligence, \bibinfo{pages}{1A1NFC1a1--1A1NFC1a1}.
\newblock


\end{thebibliography}

\end{document}